\documentclass[paperpaper, 10pt, conference]{ieeeconf}  % Comment this line out if you need a4paper

\IEEEoverridecommandlockouts                              % This command is only needed if 
\usepackage{graphicx}
\usepackage{booktabs}
\usepackage{color,array}
\usepackage{xcolor}
\usepackage[misc]{ifsym}

\usepackage{microtype}
\usepackage{subfigure} 
\usepackage[colorlinks,urlcolor=blue,linkcolor=blue,citecolor=blue]{hyperref}

\usepackage{amssymb}
\usepackage{amsmath}
\usepackage{amsfonts}
\usepackage{cite}
\usepackage{siunitx}
\usepackage{bm}
\usepackage{cancel}
\usepackage{longtable,tabularx}
\usepackage{multirow}
\usepackage{placeins}
\usepackage{float}
\usepackage{algorithm}
\usepackage{algorithmic}
\usepackage{tikz}
\usetikzlibrary{patterns}
\usepackage{mathrsfs} 

\usepackage{hyperref}

\newtheorem{condition}{Condition}
\newtheorem{definition}{Definition}
\newtheorem{theorem}{Theorem}
\newtheorem{remark}{Remark}

\def\be{\begin{equation}}
\def\ee{\end{equation}}

\newcommand{\MK}[1]{{#1}}
\newcommand{\YL}[1]{{#1}}

\newcommand{\LY}[1]{{#1}}

\newtheorem{lemma}{Lemma}

\newcommand{\RV}[1]{{\color{black} #1}}

\usepackage{mathtools}

\title{\LARGE \bf
On Space-Filling Input Design for Nonlinear Dynamic Model Learning: \\
A Gaussian Process Approach
}

\author{Yuhan Liu, Máté Kiss, Roland Tóth and Maarten Schoukens% <-this % stops a space
\thanks{This work is funded by the European Union (ERC, COMPLETE, 101075836). Views and opinions expressed are however those of the author(s) only and do not necessarily reflect those of the European Union or the European Research Council Executive Agency. Neither the European Union nor the granting authority can be held responsible for them.}% <-this % stops a space
\thanks{Y. Liu, M. Kiss, R. Tóth and M. Schoukens are with the Control Systems Group, Electrical
Engineering, Eindhoven University of Technology, 5600 MB Eindhoven, The
Netherlands. {\tt\small (email: y.liu11@tue.nl).}}%
\thanks{R. Tóth is also with the Systems and Control Laboratory, HUN-REN Institute for Computer Science and Control, 1111 Budapest, Hungary.
        }%
}

\begin{document}

\maketitle
\thispagestyle{empty}
\pagestyle{empty}

%%%%%%%%%%%%%%%%%%%%%%%%%%%%%%%%%%%%%%%%%%%%%%%%%%%%%%%%%%%%%%%%%%%%%%%%%%%%%%%%
\begin{abstract}

While optimal input design for linear systems has been well-established, no systematic approach exists for nonlinear systems where robustness to extrapolation/interpolation errors is prioritized over minimizing estimated parameter variance.
To address this issue,
\RV{we develop a novel space-filling input design strategy for nonlinear system identification that ensures data coverage of a given 
region of interest.}
By placing a Gaussian Process (GP) prior on the joint input-state space, the proposed strategy leverages the GP posterior variance to construct a cost function that promotes space-filling input design. 
Consequently, this enables maximization of the coverage in the region of interest, thereby facilitating the generation of informative datasets. Furthermore, we theoretically prove that minimization of the cost function implies the space-filling property of the obtained data. 
Effectiveness of the proposed strategy is demonstrated on both an academic and a mass-spring-damper example, highlighting its potential practical impact on efficient exploration of the dynamics of nonlinear systems.
\end{abstract}
%
% \begin{IEEEkeywords}Space filling, Input design, Gaussian process, System identification, Dynamical system. 
% \end{IEEEkeywords}

%%%%%%%%%%%%%%%%%%%%%%%%%%%%%%%%%%%%%%%%%%%%%%%%%%%%%%%%%%%%%%%%%%%%%%%%%%%%%%%%
\section{Introduction}
\label{sec:1}

\RV{When performing nonlinear system identification \cite{schoukens2019nonlinear}, the quality of the model strongly depends on the quality of the data.} 
\RV{In general, experiment design aims to obtain the information required to model system dynamics at the lowest experimental cost, while achieving expected modeling accuracy and respecting possible constraints to the experiment itself \cite{schoukens2019nonlinear,bombois2006least}.
In-depth research has been conducted on input design for \emph{linear time-invariant} (LTI) systems \cite{tanaskovic2014optimal,annergren2017application,bombois2021robust}.}  {State-of-the-art methods} minimize the variance on the parameter estimate of an assumed parametric model of the system by maximizing the Fisher information matrix $\mathcal{M}$, under nominal and robust settings by D-, A, or E-optimality criterion.
In the LTI case, this leads to convex optimization problems, giving optimal design of the power spectrum of the input signal~\cite{pronzato2008optimal}.  \RV{Other approaches focus explicitly on the intended application of the model, known as \emph{application-oriented} input design (see \cite{bombois2006least,hjalmarsson2009system,annergren2017application}, etc.), which often links the design criteria to a specific use case such as control performance.}

\RV{In contrast to the linear case, nonlinear systems generally require the input design to account for both spectral and time-domain characteristics, such as amplitude distribution and temporal evolution due to their inherently coupled behavior.}
\RV{Most existing methods target specific classes of nonlinear systems, such as Hammerstein and Wiener such as \cite{de2016d}. %or focus on statistical properties of the obtained model, like \cite{valenzuela2015graph,parsa2023coherence}}. 
Others consider a wider class of systems but are limited to an input sequence with a finite number of different values \cite{valenzuela2015graph}, or aim to minimize the mutual coherence of the regressor in the sparse nonlinear system identification problem \cite{parsa2023coherence}.}

Given that the dominant source of errors in nonlinear system identification is most often the generalization error, i.e., model error over a region of the operating space that is not contained in the estimation dataset \cite{schoukens2019nonlinear}, input design that minimizes the variance of the parameter estimate is not justified. Instead, the design should focus on minimizing the model error over an operating range of interest.

A space-filling input ensures that, after identification, the model errors are well behaved and that the estimated model generalizes well over the full region of interest.
\RV{Various methods \cite{heinz2017iterative,smits2024space,peter2019fast,KISS2024562,vater2024differentiable,herkersdorf2024optimized} have been explored for the design of space-filling inputs for nonlinear systems, utilizing, for instance, distance-based criteria that penalize mutual proximity between data points \cite{heinz2017iterative,smits2024space,peter2019fast}, as well as more recent approaches based on Jensen-Shannon divergence \cite{vater2024differentiable} or frequency-aware space-filling costs \cite{herkersdorf2024optimized}. However, most existing approaches are restricted to specific input classes (e.g., piecewise constant inputs \cite{smits2021genetic,smits2022space,smits2024space}) and do not consider constraints on the input (e.g., spectral constraints). In addition, some recent methods require matching the empirical data distribution to a fixed target distribution, which may be unnecessarily restrictive, or rely on non-smooth cost functions that limit the use of gradient-based optimization.}
Active learning, which aims at similar objectives as input design, can also be used to actively explore the space of interest by iteratively selecting sampling points that, based on the current model estimate, maximize a variance-based cost \cite{buisson2020actively} or the information gain \cite{treven2023optimistic}. 
In contrast, space-filling input design, as an experiment design method, focuses on thorough coverage of the entire space by solving an offline design problem using a prior model of the system \RV{without requiring real-time state information during data collection.}

In this paper, we study how to efficiently generate an informative dataset that can {maximize the space coverage for a fixed length of the experiment}. To this end, we propose a simple, yet effective  \emph{Gaussian process} (GP)-based space-filling input design strategy. 
The underlying philosophy of the proposed method {is to solve the filling problem by} minimizing the total posterior variance w.r.t. some user-selected anchor points distributed in a region of interest. The main contributions  are summarized as: 
\begin{itemize} 
\item[C1)]
A novel GP-based space-filling input design algorithm is presented by developing a variance-based cost function to explore a user-defined region of interest efficiently; 
\item[C2)] Rigorous theoretical analysis is provided to illustrate that the cost function minimization leads to a space-filling behavior regardless of the hyperparameter choice; 
\item[C3)] The proposed approach is not limited to a specific input class, facilitating flexible choice of parameterization and allowing for a rather wide range of input constraints.
\end{itemize} 

The paper is organized as follows. The problem formulation is described in Section \ref{sec:2} and GPs are introduced in Section \ref{sec:3}.  The proposed GP-based space-filling input design method is detailed in Section \ref{sec:4}.
Examples are given in Section \ref{sec:5}, followed by the conclusions in Section \ref{sec:6}.

\section{\YL{Problem Formulation}}
\label{sec:2}
\subsection{Data-generating system}
\label{subsection:dg}

Consider the following discrete-time state-space representation of the data-generating system: 
\be
\label{1}
\begin{aligned}
x(k+1) & =f(x(k), u(k)), \\
y(k) & =g(x(k), u(k)),
\end{aligned}
\ee
where $u(k)\in\mathcal{U}\subseteq\mathbb{R}^{n_\mathrm{u}}$, $x(k)\in\mathcal{X}\subseteq\mathbb{R}^{n_\mathrm{x}}$, and $y(k)\in\mathcal{Y}\subseteq\mathbb{R}^{n_\mathrm{y}}$ are the input, state and output signals of the system at time instant $k\in\mathbb{Z}$, respectively, and $\mathcal{Z}:\mathcal{U}\times\mathcal{X}$ denotes the joint input-state space. The smooth nonlinear functions $f(\cdot):\mathcal{Z}\to\mathcal{X}$ and $g(\cdot):\mathcal{Z}\to\mathcal{Y}$ represent the known model of the system, \RV{ with $f$ assumed to be continuously differentiable}.

\begin{condition}\cite{nijmeijer1990nonlinear}
\label{cond:controllable} 
% ({{Controllability for NL-SS system}})
\RV{The state-space representation  \eqref{1} is controllable, which indicates that for any two states $x_1,x_2\in\mathcal{X}$ there exists a %finite number of steps 
$T_{\mathrm{d}}\in\mathbb{Z}_{>0}$ and an %admissible control function 
input sequence $u:\{0,1,...,T_{\mathrm{d}}-1\}\to\mathcal{U}$
% $u:[0,T_{\mathrm{d}}]\to\mathcal{U}$ 
such that $x_2 \!=\! \phi(T_{\mathrm{d}},0,x_1,u)$, where 
$\phi(k,0,x_0,u)$ denotes the unique solution of  \eqref{1} at time step $k\geq0$ for a particular %control function 
$u$ %(\cdot)$ 
and initial condition $x(0) \!=\! x_0$.}
\end{condition}

\subsection{\YL{Space-Filling Input Design Problem}}
\label{subsection:sf}

It is worth noting that the optimal input design problem corresponds to a \emph{chicken-egg dilemma}: to optimally design an input signal that results in the informative output response w.r.t. the identification objective, perfect knowledge of the underlying system dynamics in \eqref{1} is necessary; while to completely capture the dynamics of \eqref{1}, we need a data set that is informative w.r.t. our model structure. This inherent interdependency makes it challenging to effectively address both tasks concurrently within the identification cycle. In this paper, we focus on the space-filling input design problem under a known nonlinear system model \eqref{1} regardless of whether it represents full or partial knowledge of the system.

\begin{remark}
If the model is unknown, or it is approximately known with uncertainties, an iterative procedure can be adopted: one can design an experiment on the uncertain model, or start with a generic experiment, and estimate a model from this initial dataset. Subsequently, using the refined model, the space-filling input design can be repeated, taking into account the previously measured datasets until it achieves an adequate level of modeling accuracy.
\end{remark}

Consider the set of parameterized input signals as:
\be
\label{para-u}
\mathscr{U}(\Theta):=\LY{\{u(k,\theta):\mathbb{T}\times \Theta\to\mathbb{R}^{n_\mathrm{u}}\}} %\mid \theta\in{\Theta}\}}%
\ee
\LY{where $\mathbb{T}$ is a finite interval in $\mathbb{Z}$} and \eqref{para-u} satisfies the condition:

\begin{condition}
\label{cond:u}
\LY{The partial derivatives of any signal in $\mathscr{U}(\Theta)$ w.r.t. the parameter $\theta\in{\Theta}$ exist and are bounded over $\forall k\in\mathbb{T}$,
where ${\Theta}\subseteq\mathbb{R}^{n_\mathrm{\theta}}$ is considered to be a compact set. }
% Furthermore, the length of $\mathscr{U}(\Theta)$ is finite.}
\end{condition}

Note that there are no restrictions on the type and parametrization of the input signals in $\mathscr{U}(\Theta)$. In other words,  \eqref{para-u} can represent a wide class of parametric input signals such that Condition \ref{cond:u} is satisfied. Here are two examples:
\begin{itemize}
\item Multisine signal parameterized by $\theta\triangleq\mathrm{vec}(A_l,\phi_l)\in\mathbb{R}^{n_\theta}$ with $n_\theta\!=\!2N_f$%:
\begin{small}
\be
\label{multisine}
\mathscr{U}_{\text{ms}}(\Theta)\!:=\!\left\{\! u(k,\theta)\!=\!\!\sum_{l=0}^{N\!_f\!-\!1}\!A_l\!\sin\!\left(\!2\pi l\frac{f_0}{f_s}\!k\!+\!\phi_l\!\right) \!\bigg| \theta\! \in\! \Theta \!\right\}\!
\ee
\end{small}%
where $A_l\in \mathbb{R}^{n_\mathrm{u}}$ and $\phi_l\in [0,2\pi]$ denote the amplitude and phase, respectively, \MK{$f_0$} represents the base frequency of the multisine, and $f_s$ is the sampling frequency.
\item Piecewise constant signal parameterized by  $\theta\triangleq\mathrm{vec}(A_l,p_l)\in\mathbb{R}^{n_\theta}$ with $n_\theta\!=\!2N+1$:
\be
\label{piecewise}
\mathscr{U}_{\text{pw}}(\Theta)\!:=\!\left\{\! u(k,\theta)\!=\!\!\!\sum_{l=0}^{N\!_p-1}\!A_l\delta\!\left(p_{l},p_{l+1},k\right) \!\bigg| \theta\! \in\! \Theta \!\right\}  
\ee
where $A_l\in \mathbb{R}^{n_\mathrm{u}}$ denotes the amplitude levels and 
\be
\delta(p_{l},p_{l+1},k) = \begin{cases}1, & p_{l} \leq k<p_{l+1} \\ 0, & \text{otherwise}\end{cases}
\ee denotes the duration of each amplitude level with switching instances $p_{l} \in\mathbb{T}$.
\end{itemize}

As discussed in Section \ref{sec:1}, \LY{a space-filling input allows {the model to represent the system well}} over the region of interest, which is defined~as:
\begin{definition}
\label{ROI}
A region of interest is a compact subset of the joint input-state space $\tilde{\mathcal{Z}}\subseteq\mathcal{Z}$, where the behavior of the system is of particular interest for data-driven modeling.
\end{definition}

The primary goal of this paper is to design the parameter vector $\theta\in{\Theta}$ such that when the input signal $u(k,\theta)$ is applied to the assumed, known system  \eqref{1}, the resulting dataset is expected to achieve an effective space-filling property, ensuring a good coverage of $\tilde{\mathcal{Z}}$.

\section{Gaussian Process Models}
\label{sec:3}
% \subsection{\YL{Input Design for Data-driven Model Learning}}
\LY{In this section, we introduce some basic concepts \RV{of} GP modeling}. GP regression \cite{rasmussen2003gaussian} is a well-known tool for constructing nonparametric, probabilistic models directly from data, and allows for flexible specification of prior assumptions on the system.
\LY{To construct a \emph{hypothetical} model of $f:\mathcal{Z}\to\mathcal{X}$, we consider $h:\mathcal{Z}\to\mathcal{X}$ as a GP.}
In terms of definition, a \emph{Gaussian Process} $\mathcal{GP}: \mathbb{R}^{n_\mathrm{z}} \to \mathbb{R}$ assigns to every point $ {z} \in \mathbb{R}^{n_\mathrm{z}}$ a random variable ${h}({z})$ taking values in $\mathbb{R}$ such that, for any finite set of points $\{{z}_{\tau}\}_{\tau\!=\!1}^N \subset \mathbb{R}^{n_\mathrm{z}}$, \RV{the joint probability distribution of $h\left( {z}_1\right), \ldots, h\left( {z}_{N}\right)$ is  Gaussian}. A GP can be fully characterized by its mean function $\mu(z):\mathcal{Z}\to\mathbb{R}$ and its covariance function $\kappa(z,z'):\mathcal{Z}\times\mathcal{Z}\to\mathbb{R}$ and  it is denoted as:
$h(z)\sim\mathcal{GP}\left({\mu}({z}), {\kappa}\left( {z},  {z}^{\prime}\right)\right)$.

Given a training dataset $\mathcal{D}\!=\!\{Z,H\}$ with the training input matrix $Z \!=\!\mathrm{vec}(z_1,...,z_N)\in\mathbb{R}^{N\times n_\mathrm{z}}$ and the training output $H \!=\!\mathrm{vec}(h_1,...,h_N)\in\mathbb{R}^{N\times 1}$, with $h_i\!=\!h(z_i)$. Under the assumption that the function $h(z)$ follows a GP, one can define the Gaussian prior distribution for %the function 
$h$ as $\mathbb{P}\left(H\right) \!=\! \mathcal{N}(H\mid 0,K_N)$  where  $K_N\in\mathbb{R}^{N\times N}$ is the Gram matrix with entries $[K_N]_{i,j} \!=\! \kappa(z_i,z_j)$. \RV{Given a query point $\bar{z}\in\mathbb{R}^{n_\mathrm{z}}$, let $\bar{h}\!:=\!h(\bar{z})$ denote the random output at $\bar{z}$. Then, the joint distribution of $H$ and $\bar{h}$ is}:
{\setlength\abovedisplayskip{1mm} \setlength\belowdisplayskip{1mm}
\be
\left(\begin{array}{c}
H \\ 
\bar{h}\end{array} \Bigg| \begin{array}{c}
Z \\
\bar{z}
\end{array}\right) \sim \mathcal{N}\left( \begin{array}{c}
{0}
\end{array},\begin{bmatrix}
{K}_{N} & {\kappa}(\bar{z},Z)^{\top} \\
{\kappa}(\bar{z},Z) & {\kappa}(\bar{z},\bar{z})
\end{bmatrix}\right).
\ee}%
Then conditioning the joint distribution on  $\mathcal{D}$ leads to the predictive distribution of $\bar{h}$ given an input $\bar{z}$: 
$\mathbb{P}(\bar{h}\mid\bar{z},\mathcal{D}) 
\!=\!\mathcal{N}(\bar{h}\mid \mu(\bar{h}\mid\bar{z},\mathcal{D}), \mathrm{Var}(\bar{h}\mid\bar{z},\mathcal{D}))$ with the mean and variance expressed respectively as:
{\setlength\abovedisplayskip{1mm} \setlength\belowdisplayskip{1mm}
\begin{subequations}
\begin{align}
\label{mean}\mu(\bar{h}\mid\bar{z},\mathcal{D})&\!=\! {\kappa}(\bar{z},Z)K_N^{-1}H,\\
 \label{var}
 \mathrm{Var}(\bar{h}\mid\bar{z},\mathcal{D})&\!=\!{\kappa}(\bar{z},\bar{z})\!-\!{\kappa}(\bar{z},Z)K_N^{-1}{\kappa}(\bar{z},Z)^{\top}.
 \end{align}
\end{subequations}}%

\begin{remark}
\label{remarl:variance}
According to \eqref{var}, %it can be deduced that 
the posterior variance evaluated at $\bar{z}$ only depends on the GP inputs $Z$. %which corresponds to the {particular property of Gaussian distribution}. 
Hence, the posterior variance of a GP inherently represents the distribution of GP inputs $Z$ {in the given dataset over %the space 
$\mathcal{Z}$}. 
\end{remark}

\section{GP-based Space-filling Input Design}
\label{sec:4}
\YL{In this section, we propose a GP-based space-filling input design algorithm, enabling the acquisition of an informative dataset for data-driven model learning that achieves good coverage within the region of interest $\tilde{\mathcal{Z}}$. The main philosophy behind this is to firstly place a GP prior on a latent \emph{hypothetical} model $\hat{x}(k+1)\!=\!\hat{f}(x(k),u(k))$ of the data-generating system  \eqref{1}, and then construct a space-filling promoting cost function from the Bayesian point of view. }

\subsection{Classical Space-filling Design}\label{sec:4:1}
In this subsection, we briefly introduce the concept of the classical space-filling design method \cite{johnson_minimax_1990}.

\begin{definition}
\label{SF}
Consider a dataset $\mathcal{D}_N\!=\!\{z_i\}_{i\!=\!1}^N$ and a region of interest ${\tilde{\mathcal{Z}}}$. 
Let $d(\varsigma,\mathcal{D}_N)\!=\!\min_{z_i\in\mathcal{D}_N}d(\varsigma,z_i)$ denote the minimum distance w.r.t. $\varsigma$ where $d(\cdot,\cdot):\mathcal{Z}\times\mathcal{Z}\to\mathbb{R}_0^{+}$ is a user-chosen metric (e.g. Euclidean distance).  Then
{\setlength\abovedisplayskip{1pt} \setlength\belowdisplayskip{1pt}
\be
\LY{\rho({\mathcal{D}_N})} = \max_{\varsigma\in\tilde{\mathcal{Z}}}d(\varsigma,\mathcal{D}_N)
\ee}% 
is the so-called filling distance indicating the radius of the largest ball that can be placed in $\tilde{\mathcal{Z}}$ which does not contain any point in $\mathcal{D}_N$.
Specifically, if there exists a constant $\epsilon$ such that $\LY{%\min_{\mathcal{D}_N}
\rho({\mathcal{D}_N})}<\epsilon, ~\epsilon>0$, then \LY{$\mathcal{D}_N$ is said to have the at least $1/\epsilon$ density w.r.t $\tilde{\mathcal{Z}}$, i.e., the $\epsilon$-space filling property.}
\end{definition}

In contrast to this work, classical space-filling design only focuses on the distribution of the points within $\mathcal{D}_N$, without considering the temporal dynamics between the points, i.e., the underlying dynamics that generate these data points.

\subsection{Space-filling Promoting Cost Function}
The proposed space-filling promoting cost function is constructed based on the dataset $\mathcal{D}_N\!=\!\{z_j\}_{j\!=\!1}^{N}$ generated by the parameterized input signal $u(k,\theta)$ after applying it to the \RV{known model} of the system  \eqref{1}. Based on the dataset $\mathcal{D}_N$, one has following data matrix $Z \!=\! \mathrm{vec}(z_1,\ldots,z_N)\in\mathbb{R}^{N\times n_z}$ 
where $z_i \!=\! \mathrm{vec}(x(i),u(i,\theta))$, $i\in\mathbb{I}_{i\!=\!1}^{N}$.

Additionally, to facilitate the design of the cost function, we define anchor points as follows in our learning scenario:
\begin{definition}
\label{def:inf}
Given a region of interest $\tilde{\mathcal{Z}}\subseteq\mathcal{Z}$, the {anchor dataset} is characterized by $\mathcal{D}_\mathrm{I}\!:=\!\{\tilde{z}_i\!=\!\mathrm{vec}(\tilde{x}_i,\tilde{u}_i)\}_{i\!=\!1}^{M}$, which contains $M$ pre-defined anchor points from $\mathcal{Z}$ %to capture the dynamics, thereby 
gridding the region of interest~$\tilde{\mathcal{Z}}$.
\end{definition}

\RV{We place a GP prior on the latent \emph{hypothetical} model $\hat{x}(k+1)\!=\!\hat{f}(x(k),u(k))$ of the known data-generating system  \eqref{1} which results in $\hat{f}_\nu\sim\mathcal{GP}\left({\mu}_\nu({z}), {\kappa}\left( {z},  {z}^{\prime}\right)\right), \nu\in\mathbb{I}_1^{n_\mathrm{x}}$, where all $\hat{f}_\nu$ share the same kernel ${\kappa}\left( {z}, {z}^{\prime}\right)$ and input space. This
constitutes to $\hat{f} \!=\!\mathrm{vec}(\hat{f}_1,...\hat{f}_{n_\mathrm{x}})$}.
The kernel function $\kappa\left({z},{z}^{\prime}\right)$ satisfies the following assumption:
\begin{condition}
\label{assump:kernel}
\RV{The kernel function $\kappa(z,z')\!=\!\kappa(\|z-z'\|)$ is chosen to be a positive definite kernel that is monotonically decreasing %$\tilde\kappa$ 
w.r.t. $\|z-z'\|$.}
\end{condition}

The decreasing monotonicity of $\kappa(z,z')$ indicates that the association between the anchor points and $Z$ diminishes with the increase of the Euclidean distance $\|z-z'\|$.
The kernel is 
usually chosen from the exponential family, e.g., the \emph{Squared Exponential Automatic Relevance Determination} (SEARD) 
$\kappa\left({ {z}},  {  {z}}'\right)\!=\!\sigma_{\mathrm{f}}^{2} \exp \left(-\frac{1}{2}( {  {z}}\!-\!  {  {z}}')^{\top} {\Lambda}^{-1}( {  {z}}\!-\! {  {z}}')\right)\!$ is taken as a prior with $\sigma_{\mathrm{f}}^{2}\in\mathbb{R}^{+}$ and~$\Lambda \!=\! \mathrm{diag}(\ell_1^2,...,\ell_{n_\mathrm{z}}^2)\in\mathbb{R}^{n_\mathrm{z}\times n_\mathrm{z}}$, \RV{where both $\sigma_{\mathrm{f}}^{2}$ and $\Lambda$ are treated as hyperparameters.}

\RV{As all $\hat{f}_\nu$ share the same kernel and input space, they yield identical posterior variance  $\forall \nu\in\mathbb{I}_1^{n_\mathrm{x}}$. 
For simplicity, we drop the dimension indexing and refer to $\mathrm{Var}(\hat{f}\mid\tilde{z}_i,\mathcal{D}_N)$ instead of $\mathrm{Var}(\hat{f}_\nu\mid\tilde{z}_i,\mathcal{D}_N)$ for the remainder of the paper.}

To obtain a good coverage for the region of interest, the posterior variance at each anchor point:
{\setlength\abovedisplayskip{1mm} \setlength\belowdisplayskip{1mm}
\be
\label{varcost}
\mathrm{Var}(\hat{f}\mid\tilde{z}_i,\mathcal{D}_N)\!=\!{\kappa}(\tilde{z}_i,\tilde{z}_i)-{\kappa}(\tilde{z}_i,Z)K_N^{-1}{\kappa}(\tilde{z}_i,Z)^{\top}
\ee}%
should be minimized. This leads to optimizing the scalar-valued space-filling promoting cost function defined~as:
{\setlength\abovedisplayskip{1pt} \setlength\belowdisplayskip{1pt}
\be
\label{GPcost-data}
\mathcal{V}(\mathcal{D}_N):=\frac{1}{M} \sum_{i=1}^{M} \mathrm{Var}(\hat{f}\mid\tilde{z}_i,\mathcal{D}_N)
\ee}%
based on the dataset $\mathcal{D}_N$.

We would like to find a parametrized input signal $u(k,\theta)$ such that the resulting dataset $\mathcal{D}_N(\theta)$ \LY{is expected to achieve the $1/\epsilon$ density}, ensuring a good coverage within the region of interest $\tilde{\mathcal{Z}}$. To this end, when taking the dynamical system \eqref{1} into account, the posterior variance at each anchor point can be rewritten in a $\theta$-dependent form:
{\setlength\abovedisplayskip{1mm} \setlength\belowdisplayskip{1mm}
\be
\label{varcost-theta}
\mathrm{Var}(\hat{f}\!\mid\!\tilde{z}_i,\!\mathcal{D}_N\!(\theta))\!=\!{\kappa}(\tilde{z}_i,\!\tilde{z}_i)\!-\!{\kappa}(\tilde{z}_i,\!Z_\theta)K_N^{-1}{\kappa}(\tilde{z}_i,\!Z_\theta)^{\top}.
\ee}%
The space-filling input design problem can be formalized~as:
\vspace{-2mm}
{\setlength\abovedisplayskip{1pt} \setlength\belowdisplayskip{1pt}
\begin{subequations}
\begin{align}
\underset{\theta\in{\Theta}}{\text{min}} 
 & \quad \mathcal{W}({\theta};\mathcal{D}_N(\theta)) \label{f1}\\
 \text{s.t.} & \quad x(0)=x_0, \quad u(0)=u_0,\\
 & \quad x(j+1) \!=\! f(x(j),{u}(j;\theta)),\\
 &\quad \mathcal{D}_N(\theta)\!:=\!\{z_j\!=\!\mathrm{vec}(x(j),u(j,\theta))\}_{j=1}^{N},
\end{align}
\end{subequations}}%
with $Z_\theta \!=\!\mathrm{vec}(z_1,...,z_N)$, where \vspace{-0.1mm}
{\setlength\abovedisplayskip{1pt} \setlength\belowdisplayskip{1pt}
\be
\label{GPcost-data-input}
\mathcal{W}(\theta;\mathcal{D}_N(\theta)):=\frac{1}{M} \sum_{i=1}^{M} \mathrm{Var}(\hat{f}\mid\tilde{z}_i,\mathcal{D}_N(\theta)).
\ee}%

The implementation of the proposed method is summarized in Algorithm \ref{alg:1}. \RV{The computational complexity per optimization step is dominated by the posterior variance calculation which scales as $\mathcal{O}(N^3+N^2(n_{\mathrm{x}}+n_{\mathrm{u}}+M))$, and is primarily determined by the input length $N$. As the proposed method is performed offline, the computational cost remains acceptable even for moderately large $N$ and system dimensions.}

\begin{remark}
\RV{Compared to recent works \cite{vater2024differentiable,herkersdorf2024optimized}, the proposed  GP-based cost does not rely on matching a predefined distribution. Matching a specific distribution can be restrictive as the desire to cover a space does not impose a specific distribution over that space. Hence, leaving room to stay closer to the natural preferences of the system dynamics by not specifying such a target distribution is beneficial to reach the space covering goal. Furthermore, the proposed GP-based cost remains smooth and differentiable, enabling gradient-based optimization under arbitrary input parameterizations.}
\end{remark}

\begin{algorithm}[tb]
   \caption{\YL{GP-based Space-Filling Input Design for Nonlinear Dynamic Model Learning}}
   \label{alg:1}
\begin{algorithmic}
   \STATE{\bf{Step 1:}} \YL{Select the number of samples $N$, step size $\alpha$ and threshold $\delta$. Set the iteration index $\iota\!=\!0$. Initialize $u(k,\theta^{(0)})$ with $\theta^{(0)}\!=\!\theta_0$};
   \STATE{\bfseries Step 2:}  Select the region of interest $\tilde{\mathcal{Z}}$ together with the \YL{anchor dataset $\mathcal{D}_\mathrm{I}$};
   \STATE {\bfseries Step 3:} \YL{Set the GP kernel $\kappa(z,z')$ with pre-defined hyperparameters $\Lambda$, $\sigma_{\mathrm{f}}^{2}$};
   % on $\tilde{\mathcal{Z}}$
   \STATE {\bfseries Step 4:}
   \YL{\WHILE{$\|\theta^{(\iota+1)}-\theta^{(\iota)}\|\geq\delta$}
   \STATE Apply $u(k,\theta^{(\iota)})$ on  \eqref{1} and obtain $\mathcal{D}_N(\theta^{(\iota)})$;
    \FOR{$i=1$ {\bfseries to} $M$}
   \STATE Compute $\mathrm{Var}(\hat{f}~|~\tilde{z}_i,\mathcal{D}_N(\theta^{(\iota)}))$ according to  \eqref{varcost-theta};
   \ENDFOR
   \STATE Compute $\mathcal{W}(\theta^{(\iota)};\mathcal{D}_N(\theta^{(\iota)}))$ according to  \eqref{GPcost-data-input};
   \STATE Set $\theta^{(\iota+1)}=\theta^{(\iota)}-\alpha\nabla_{\theta}\mathcal{W}(\theta^{(\iota)};\mathcal{D}_N(\theta^{(\iota)}))$;
   % Solve $\theta^{(\iota+1)}=\mathrm{argmin}_{ \theta\in\Theta}~ \mathcal{W}(\theta^{(\iota)};\mathcal{D}_N)$;
   \STATE Set $\iota\gets\iota+1$;
    \ENDWHILE}

   \YL{\STATE {\bfseries Step 5:} Set $\hat{\theta}\gets\theta^{(\iota)}$, and adopt $u(k,\hat{\theta})$ as the space-filling input signal.}

\end{algorithmic}
\end{algorithm}

\vspace{-1mm}
\subsection{Theoretical Analysis}

We first illustrate the space-filling property of  $\mathcal{D}_N$ w.r.t. a space-filling anchor dataset $\mathcal{D}_\mathrm{I}$  by selecting a proper kernel function, which is stated in the following lemma.

\begin{lemma}
\label{lemma:1}
Consider a dataset $\mathcal{D}_N$ and a region of interest $\tilde{\mathcal{Z}}$ which is characterized by $M$ pre-defined anchor points  $\mathcal{D}_\mathrm{I}\!=\!\{\tilde{z}_i\}_{i\!=\!1}^{M}$.
The GP-based space-filling promoting cost function is appropriately defined as \eqref{GPcost-data}, {with a kernel function $\kappa(z,z')$ satisfying Condition \ref{assump:kernel}}. If:
\begin{itemize}
\item $\mathcal{V}(\mathcal{D}_N)\!=\!0$; 
\item \LY{The anchor dataset $\mathcal{D}_\mathrm{I}$ has   $1/\epsilon$ density;}
\item \LY{The size $N$ of dataset $\mathcal{D}_N$ is finite;}
\end{itemize}
then $\mathcal{D}_N$  \LY{is guaranteed to have the at least $1/\epsilon$ density w.r.t.~$\tilde{\mathcal{Z}}$}.
\end{lemma}

\begin{proof}
First we prove that the posterior variance at $\tilde{z}_i$: $\mathrm{Var}(\hat{f} \mid \tilde{z}_i,\!\mathcal{D}_N)\!=\!{\kappa}(\tilde{z}_i,\tilde{z}_i)\!-\!{\kappa}(\tilde{z}_i,\!Z)K_N^{-1}{\kappa}(\tilde{z}_i,Z)^{\top} \!=\! 0$ \emph{if and only if} there exists a $j\in\mathbb{I}_1^{N}$ such that  $\tilde{z}_i \!=\! z^{(j)}$.

(\emph{Sufficiency}) Assume that $\tilde{z}_i \in \mathcal{D}_N$, i.e., there exists $j\in\mathbb{I}_1^{N}$ such that  $\tilde{z}_i \!=\! z^{(j)}$. This indicates that ${\kappa}(\tilde{z}_i,Z)\!=\!{\kappa}(z^{(j)},Z)$ is the $j^{\mathrm{th}}$ row of the Gram matrix $K_N$, that is,
${\kappa}(\tilde{z}_i,Z) \!=\! \mathrm{e}_jK_N$ 
where $ \mathrm{e}_j \!=\! ( \delta_{1,j}, \dots, \delta_{N,j} )$ is a standard basis of $\mathbb{R}^{N}$ and 
$\delta_{j,i} \!=\! 1$ if  $j \!=\! i$,  and  $\delta_{j,i} \!=\! 0$ if  $j \neq i$.
Then we have: $\mathrm{Var}(\hat{f}\mid\tilde{z}_i,\mathcal{D}_N)\!=\!\kappa(z^{(j)},z^{(j)})-\mathrm{e}_jK_NK_N^{-1}K_N^{\top}\mathrm{e}_j^{\top}
\!=\!\kappa(z^{(j)},z^{(j)})-\mathrm{e}_jK_N\mathrm{e}_j^{\top}
\!=\!\kappa(z^{(j)},z^{(j)})- \kappa(z^{(j)},z^{(j)})\!=\!0$, where $\mathrm{e}_jK_N\mathrm{e}_j^{\top}$ denotes the $j^{\mathrm{th}}$ diagonal element of $K_N$. This constitutes the proof of the sufficiency.

(\emph{Necessity}) We show the necessity part by contradiction. Suppose that
$\tilde{z}_i \notin \mathcal{D}_N$,
and ${\kappa}(\tilde{z}_i,\tilde{z}_i)-{\kappa}(\tilde{z}_i,Z)K_N^{-1}{\kappa}(\tilde{z}_i,Z)^{\top}\!=\!0$ still holds. \LY{Since the kernel function $\kappa(\cdot,\cdot)$ is positive definite, the augmented matrix  $\left[\begin{smallmatrix} K_N & \kappa(\tilde{z}_i, Z)^{\top} \\ \kappa(\tilde{z}_i, Z) & \kappa(\tilde{z}_i, \tilde{z}_i) \end{smallmatrix}\right] \succ 0$
for all $\tilde{z}_i \notin \mathcal{D}_N$. As $K_N$ is positive definite, by the Schur complement, ${\kappa}(\tilde{z}_i,\tilde{z}_i)-{\kappa}(\tilde{z}_i,Z)K_N^{-1}{\kappa}(\tilde{z}_i,Z)^{\top}>0$, $\forall \tilde{z}_i \notin \mathcal{D}_N$.} This contradicts with the assumption, hence implying that only if $\tilde{z}_i \in \mathcal{D}_N$,
$\mathrm{Var}(\hat{f}\mid\tilde{z}_i,\mathcal{D}_N)\!=\!0$ holds.

Next, based on the results above, we need to further prove that the cost $\mathcal{V}(\mathcal{D}_N)\!=\!0$ \emph{if and only if} every anchor point has data on it.

(\emph{Sufficiency}) Assume that $\forall i\in\mathbb{I}_1^{M}$, the anchor point $\tilde{z}_i$
 has at least one data $z^{(j)}$ on it. Based on the proof in the first step, we have
$ \mathrm{Var}(\hat{f}\mid\tilde{z}_i,\mathcal{D}_N)\!=\!0$.
This leads to $\mathcal{V}(\mathcal{D}_N)\!=\!0$.

(\emph{Necessity}) Suppose that $\mathcal{V}(\mathcal{D}_N)\!=\!0$. According to its definition given in \eqref{GPcost-data}, it can be easily derived that
$\frac{1}{M} \sum_{i\!=\!1}^{M} \mathrm{Var}(\hat{f}\mid\tilde{z}_i,\mathcal{D}_N)\!=\!0$.
The non-negative posterior variance implies that $\mathrm{Var}(\hat{f}\mid\tilde{z}_i,\mathcal{D}_N)\!=\!0$ for all $i\in\mathbb{I}_1^{M}$. Again, based on the proof in the first step, $\mathrm{Var}(\hat{f}\mid\tilde{z}_i,\mathcal{D}_N)\!=\!0$ holds if and only if $\tilde{z}_i \in \mathcal{D}_N$. Therefore, every anchor point $\tilde{z}_i$ must be a point in $\mathcal{D}_N$, i.e., 
$\exists \, z^{(j)} \in Z$  such that  $z^{(j)} \!=\! \tilde{z}_i$.
Since the anchor dataset $\mathcal{D}_\mathrm{I}$ is designed to be space-filling inside the entire region of interest $\tilde{\mathcal{Z}}$, the dataset $\mathcal{D}_N$ must cover $\tilde{\mathcal{Z}}$ as well. This completes the proof.
\end{proof}

Based on Lemma~\ref{lemma:1}, subsequently, we show the $\epsilon$-space filling property of a dataset $\mathcal{D}_N(\theta)$ which is generated by the optimized input applied to the dynamical system. \LY{To achieve this, the following condition is assumed to be satisfied for a set of parametrized input signals $\mathscr{U}(\Theta)$ given in \eqref{para-u}:}
\begin{condition}
\label{cond:4}
\RV{There exists a sufficiently flexible parametrization for $\mathscr{U}(\Theta)$ such that for each time step $k$, the input value $u(k)$ can be freely specified without constraints by appropriately choosing $\theta$, 
for example, $u(k,\theta)\!=\!\theta_k, \forall k\in\mathbb{Z}$.}
\end{condition}

\begin{figure*}[h]
\centering 
\includegraphics[width=5.3in]{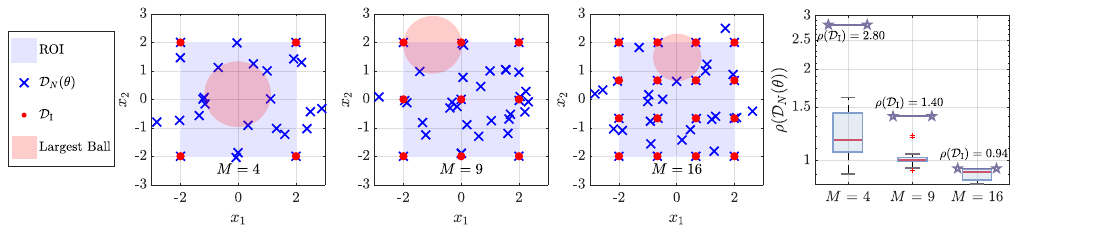}
\vspace{-4mm}
\caption{Space-filling performance of $\mathcal{D}_N(\theta)$ (\protect\tikz[baseline=-0.5ex] \protect\draw[line width=0.3mm, blue] (-0.08,-0.08) -- (0.08,0.08) (-0.08,0.08) -- (0.08,-0.08);) under different numbers of anchor points (\protect\tikz[baseline=-0.5ex] \protect\draw[red, fill=red] (0,0) circle (0.03cm);) $M=4,9,16$ with $N=40$ when the optimized $\mathcal{W}({\theta};\mathcal{D}_N(\theta))\to 0$, where \protect\tikz[baseline=-0.5ex] \protect\draw[fill=red!30, draw=red!30] (-0.15, -0.08) rectangle (0.15, 0.08); represents the largest ball in $\tilde{\mathcal{Z}}$ that does not contain any point in $\mathcal{D}_N(\theta)$. The boxplot illustrates that $\mathcal{D}_N(\theta)$ always exhibits a lower $\rho$ than the theoretically guaranteed value ($\epsilon \!=\! 2.8, 1.4, 0.94)$ over 20 Monte Carlo realizations with $\theta_k^{(0)}\sim\mathcal{N}(0,1)$, $\forall k\in\mathbb{I}_1^{N}$. \vspace{-3mm}
}\label{fig:LTI}  
\end{figure*}

Then the main theorem is stated as follows:
\begin{theorem}
\label{theo}
Consider system \eqref{1} which satisfies Condition \ref{cond:controllable} and the parameterized input signal $u(k,{\theta})$ which satisfies Conditions \ref{cond:u} and \ref{cond:4}. The GP-based space-filling promoting cost function is defined as \eqref{GPcost-data-input} \YL{on the basis of the $\epsilon$-space filling anchor dataset $\mathcal{D}_\mathrm{I}\!=\!\{\tilde{z}_i\}_{i\!=\!1}^{M}$}, with a kernel function $\kappa(z,z')$ satisfying Condition \ref{assump:kernel}. Then, the dataset $\mathcal{D}_N(\hat{\theta})$ generated by the optimized input signal $u(k,\hat{\theta})$ achieves the \LY{$\epsilon$-space filling} property in the region of interest $\tilde{\mathcal{Z}}$ with a finite data length $N\geq M \times (T_{\mathrm{d}}+1)$, where $\hat{\theta}\!=\!\mathrm{arg\,min}_{\theta\in\Theta}~ \mathcal{W}(\theta;\mathcal{D}_N(\theta)).$
\end{theorem}

\begin{proof}
On the basis of Condition \ref{cond:controllable}, it is obvious that starting from  any $z_j\!=\!\mathrm{vec}(x(j),u(j))\in\mathcal{Z}$, it takes 1 step to reach $x(j+1)$, and another $T_\mathrm{d}$ steps to reach any anchor point $\tilde{z}_i\!=\!\mathrm{vec}(\tilde{x}(i),\tilde{u}(i))\in\tilde{\mathcal{Z}}$, $\forall i\in\mathbb{I}_1^M,j\in\mathbb{I}_1^N$.
Hence, all $M$ anchor points can be explored with a finite data length $N\geq M \times (T_{\mathrm{d}}+1)$.
Condition \ref{cond:4} guarantees that the parametrization of $\mathscr{U}({\Theta})$ is sufficiently flexible to generate any value of the input. 
Combining again with Condition \ref{cond:controllable}, it can be deduced that there exists a parameter vector $\theta\in\Theta$ such that the corresponding input signal $u(k,{\theta})$ can generate a dataset $\mathcal{D}_N(\theta)$ that satisfies $\mathcal{V}(\mathcal{D}_N)\!=\!0$.
Then according to Lemma \ref{lemma:1}, one can readily conclude that the dataset $\mathcal{D}_N(\hat{\theta})$  generated by  $u(k,\hat{\theta})$ optimized based on the cost function \eqref{GPcost-data-input} \LY{is guaranteed to have the $1/\epsilon$ density w.r.t. $\tilde{\mathcal{Z}}$.} 
\end{proof}

\begin{remark}
Theorem \ref{theo} implies that minimizing \eqref{GPcost-data-input} leads to the $\epsilon$-space filling regardless of the hyperparameter choice. It is worth mentioning that our proposed space-filling input design strategy is not concerned with the GP mean estimate, as the standard GP \emph{regression} does. In practice, the selection of $\sigma_{\mathrm{f}}^{2}$ and $\Lambda$ can make the problem easier to optimize.
Moreover, in the finite data case, the length scale $\Lambda$ should be chosen considering both the interval between anchor points and the size  $N$ to ensure sufficient coverage.
\end{remark}

\vspace{-1mm}
\section{\YL{Simulation Experiments}}
\label{sec:5}
In this section, simulation results for a linear system and a nonlinear mass-spring-damper system are presented to demonstrate the effectiveness of our proposed strategy. 
Note that to quantify the space-filling behavior of the proposed algorithm, we employ the distance-based metric presented in Section \ref{sec:4:1} to verify whether the dataset $\mathcal{D}_N(\theta)$ also exhibits classical space-filling properties. The criterion $\rho$  w.r.t. a dataset $\mathcal{D}$ is denoted~as:
{\setlength\abovedisplayskip{1pt}
\setlength\belowdisplayskip{1pt}
\be
\label{rho-metric}
\rho(\mathcal{D}) = \max_{\varsigma\in\tilde{\mathcal{Z}}}\min_{z_j\in\mathcal{D}}\|\varsigma-z_j\|_{Q},~\quad \forall  j\in\mathbb{I}_1^{|\mathcal{D}|},
\ee}%
where $\varsigma$ represents a set of uniformly gridded points over $\tilde{\mathcal{Z}}$, with 100 points along each dimension, $\|\varsigma-z_j\|_{Q}\!=\!\sqrt{(\varsigma-z_j)^{\top}Q(\varsigma-z_j)}$
denotes the normalized distance between $\varsigma$ and $z_j$.
\vspace{-2mm}
\subsection{Linear system example}
Consider the continuous-time linear time-invariant system:
{\setlength\abovedisplayskip{1pt} \setlength\belowdisplayskip{1pt}
\be
\label{16}
\dot{x} = Ax + Bu, \quad A\!=\!\begin{bmatrix}
0 & 1 \\
-0.3 & -0.5
\end{bmatrix}, \quad B\!=\!\begin{bmatrix}0 \\
1
\end{bmatrix}
\ee}%
Model \eqref{16} is discretized using zero-order hold on the input with a sample time of $T_s \!=\! 1$s. The region of interest is selected as  $\tilde{\mathcal{Z}} \!=\! [-2,2] \times [-2,2]$. The input sequence is a fully parameterized signal $u(k,\theta)\!=\!\theta_k$ with  $N\!=\!40$ samples, where the input value at each time step is treated as an optimization variable. Initially, $\theta_k^{(0)}\sim\mathcal{N}(0,1)$ for all $k\in\mathbb{I}_1^{N}$.
Figure~\ref{fig:LTI} illustrates the space-filling performance under different numbers of anchor points when the optimized cost tends to 0 with $Q$ set as an identity matrix. 
It can be observed that $\mathcal{D}_N(\theta)$ achieves a lower $\rho$ than the theoretically guaranteed space-filling bound ($\epsilon \!=\! 2.8, 1.4, 0.94$) over the 20 Monte Carlo realizations.
Moreover, as  $M$ increases, the radius of the largest ball decreases, indicating a corresponding increase in density, thereby leading to improved space-filling performance.
\vspace{-1mm}
\subsection{Nonlinear mass-spring-damper system example}
\RV{This example illustrates the effectiveness of the proposed experiment design approach for nonlinear systems, even when Condition~\ref{cond:4} is violated by limiting the frequency and amplitude spectrum range of the input signal.}

We consider a nonlinear mass-spring-damper which consists of a horizontally moving mass fixed in a rail connected by a spring to the ceiling. The system is excited by an external force $F$.
By first-principle modeling, one can derive the dynamics in terms of state-space representation \cite{KISS2024562}:
{\setlength\abovedisplayskip{1mm} \setlength\belowdisplayskip{1mm}
\begin{equation}
\label{MSD}
\dot{x}_1\!=\! x_2,
\dot{x}_2 \!=\! \left(F\! - \!b\eta^{-1}(x_1)\!\left(\alpha(x_1) - l\right){x_1} \!-\! c x_2\right)/m
\end{equation}}%
where $x_1$ and $x_2$ denote the position and velocity of the moving body, respectively, and $\eta(x_1)\!=\!\sqrt{x_1^2+a^2}$. The physical parameters are set as $l\!=\!0.17$~m, $a\!=\!0.25$~m, $m\!=\!5$~kg,  $b\!=\!800$~N/m, and  $c\!=\!10$~Ns/m.

A multisine excitation (\ref{multisine}) is considered with $N_f \!=\! 92$ excited frequencies between $[1,10]$ Hz, a sampling frequency $f_s \!=\! 100$ Hz, $N \!=\! 1024$ points per period, and with $k_{\mathrm{min}}\!=\! 11$ and $k_{\mathrm{max}} \!=\! 102$. 
Both the amplitudes and the phases of each frequency component are optimized, denoted as $\theta \!=\! \mathrm{vec}(A_l,\phi_l),l\in\mathbb{I}_{0}^{N_f-1}$.
The phase $\phi_l$ is initialized by sampling from a uniform distribution $\mathcal{U}(0,2\pi)$, and the initial amplitude value $A_l$ is selected as 100 N. In addition, to limit the input signal power, we set the upper bound of $A_l$ to 200 N. To simplify the optimization problem, we assume that all frequency components share the same amplitude. The region of interest is selected as $\tilde{\mathcal{Z}} \!:= \! [-400,400] \!\times\! [-2,2] \!\times\! [-20,20]$, with  $M \!=\! 512$ uniformly distributed anchor points $\tilde{z}_i$ with $i\in\mathbb{I}_{1}^{M}$. The kernel hyperparameters are chosen as $\sigma_{\mathrm{f}}^{2}\!=\!10$ and  $\Lambda^{1/2} \!=\! \mathrm{diag}(120, 0.6, 6)$.

\begin{figure}[t]
\centering     
\includegraphics[width=2.2in]{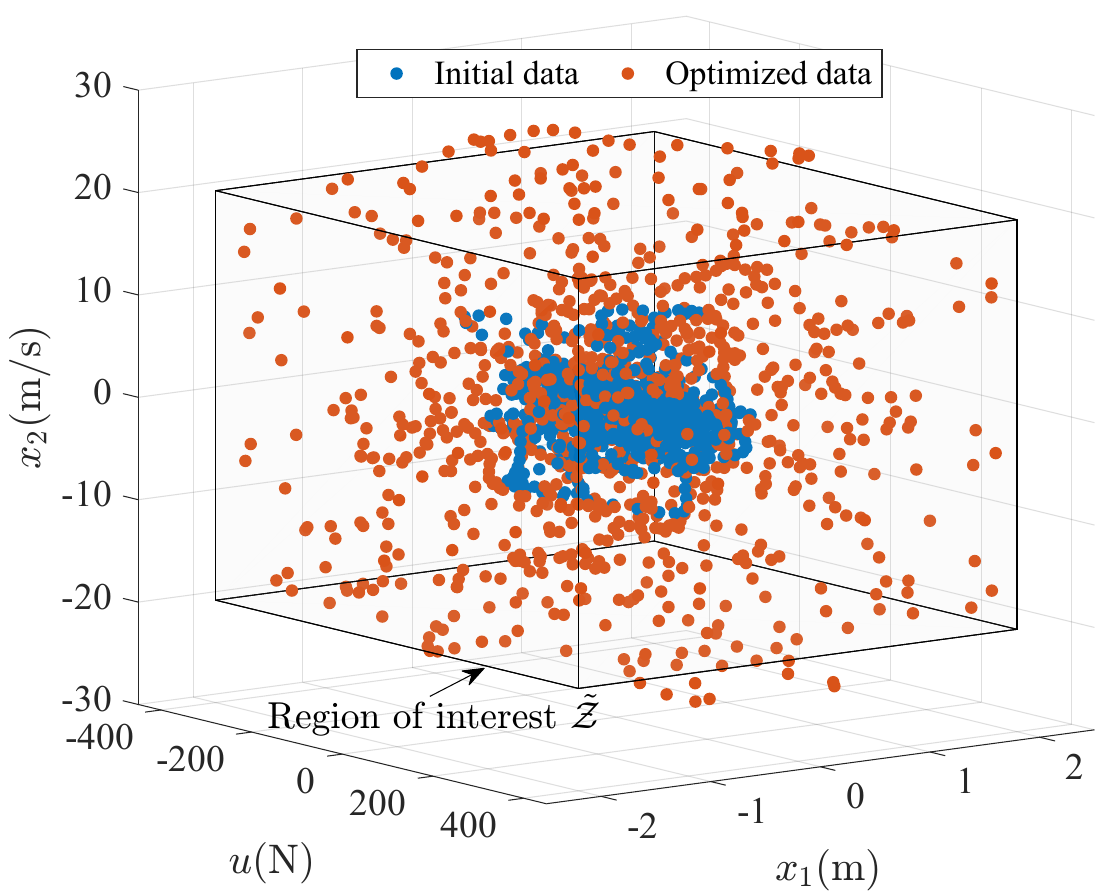}  
% 2D_isometricView
\vspace{-3mm}
\caption{{Comparison of initial dataset $\mathcal{D}_N(\theta^{0})$(\protect\tikz \protect\draw[line width=0.8mm, color={rgb:red,0;green,0.43;blue,0.75}] (0,0) -- (0.08,0);) and optimized points $\mathcal{D}_N(\hat{\theta}) $(\protect\tikz \protect\draw[line width=0.8mm, color={rgb:red,0.85;green,0.35;blue,0.13}] (0,0) -- (0.08,0);) in the 3D space. The anchor points are shown by the shaded~cube.}}
\label{fig:isometricView} %\vspace{-3mm}
\end{figure}

A 3D depiction of the considered domain is shown in Fig.~\ref{fig:isometricView}. Although the initial data is highly concentrated around the center of the domain, after optimization, a clear space-filling behavior can be observed. 
It is important to emphasize that these results were observed using a $N \!=\! 1024$ while exciting only a limited frequency range. 
Table~\ref{tab:1} provides a quantitative evaluation of $\rho(\mathcal{D}_N(\theta))$ and $\rho(\mathcal{D}_\mathrm{I})$ (mean value over 10 Monte Carlo simulations) under different sizes of $\mathcal{D}_\mathrm{I}$ before and after optimization with the normalization weight $Q \!=\! \mathrm{diag}\left(\frac{1}{400^2},\frac{1}{2^2},\frac{1}{20^2}\right)$.
 It can be observed that minimizing the proposed GP-based space-filling promoting cost function  \eqref{GPcost-data-input} leads to the minimization of the classical criterion \eqref{rho-metric}.  \RV{As the input is restricted to a user-defined frequency range and amplitude spectrum, we expect local minima in the proposed cost due to the nonlinear system dynamics, a nonzero optimal cost can be expected. While $\rho(\mathcal{D}_N(\theta))$  is slightly larger than  $\rho(\mathcal{D}_\mathrm{I})$  in the $M\!=\!512$ case, the method still significantly improves the space-filling property of the dataset over the unoptimized input, validating its robustness even beyond the strict theoretical conditions.}
 \RV{We further compare the proposed method with a Schroeder phase multisine, a common choice in experimental design. The corresponding $\rho(\mathcal{D}_{N}^{\text{MS}}) \!=\! 1.1811$, which is lower than the unoptimized input but significantly higher than the values obtained by the proposed approach. This demonstrates the improvement achieved by the proposed strategy over conventional input design choices.
}

\begin{figure}[t]
\centering     
\includegraphics[width=1.15in]{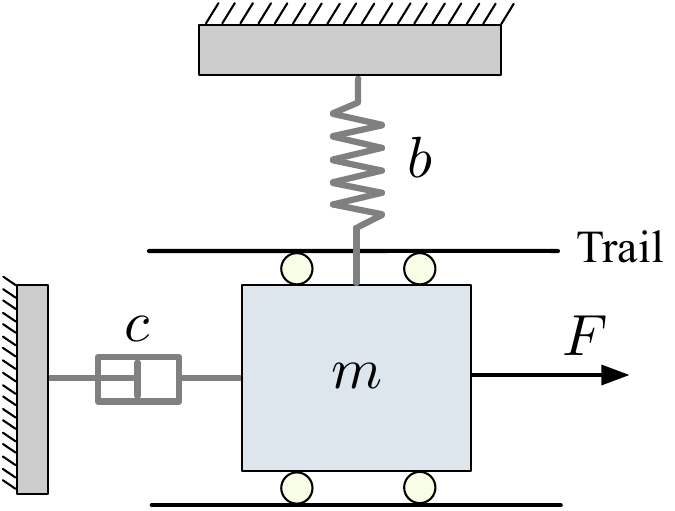} 
\caption{An illustration of the nonlinear mass-spring-damper system.} \vspace{-1mm}
\label{fig:4}
\end{figure}

\begin{table}[t]
        \centering
        \caption{\RV{Quantitative evaluation of $\rho$ under different sizes of $\mathcal{D}_\mathrm{I}$}}
        \vspace{-3mm}
        \label{tab:1}
        \renewcommand\arraystretch{1.2}
       \begin{tabular}{ccccc}
        \toprule[1.5pt]
        \multicolumn{2}{c}{Num. of anchor points} & $\rho(\mathcal{D}_N(\theta))$ & $\rho(\mathcal{D}_\mathrm{I})$ & $\rho(\mathcal{D}_{N}^{\text{MS}})$         \\ \hline
        \multirow{2}{*}{$M=512$}      & Init.     & 1.2756               & \multirow{2}{*}{0.2449} & \multirow{6}{*}{1.1811} \\ \cline{2-2}
                                                                  & Opt.      &  \bf{0.3029}              &                    \\ \cline{1-4}
        \multirow{2}{*}{$M=216$}      & Init.     & 1.2862             & \multirow{2}{*}{0.3429} \\ \cline{2-2}
                                                                  & Opt.      & \bf{0.3229}           &                    \\ \cline{1-4}
        \multirow{2}{*}{$M=125$}      & Init.     & 1.2816            & \multirow{2}{*}{0.4286} \\ \cline{2-2}
                                                                  & Opt.      & \bf{0.3500}              &       \\ \bottomrule[1.2pt]             
        \end{tabular}
        
        \end{table}
        \vspace{-1mm}

\vspace{-1mm}
\section{Conclusion}
\label{sec:6}

This paper proposes an effective GP-based space-filling input design strategy for nonlinear system identification. A GP prior is placed on the latent joint input-state space, resulting in a novel space-filling promoting cost function from a Bayesian perspective under flexible parameterization of the input signal. Theoretical analysis demonstrates the equivalence between the convergence of the cost function and the expected space-filling behavior.
Effectiveness of the proposed strategy has 
been analyzed and demonstrated on experiment design examples, showing the space-covering nature of the obtained data.

\vspace{-2mm}

\bibliographystyle{IEEEtran}
\bibliography{references}

\begin{thebibliography}{10}
\providecommand{\url}[1]{#1}
\csname url@rmstyle\endcsname
\providecommand{\newblock}{\relax}
\providecommand{\bibinfo}[2]{#2}
\providecommand\BIBentrySTDinterwordspacing{\spaceskip=0pt\relax}
\providecommand\BIBentryALTinterwordstretchfactor{4}
\providecommand\BIBentryALTinterwordspacing{\spaceskip=\fontdimen2\font plus
\BIBentryALTinterwordstretchfactor\fontdimen3\font minus
  \fontdimen4\font\relax}
\providecommand\BIBforeignlanguage[2]{{%
\expandafter\ifx\csname l@#1\endcsname\relax
\typeout{** WARNING: IEEEtran.bst: No hyphenation pattern has been}%
\typeout{** loaded for the language `#1'. Using the pattern for}%
\typeout{** the default language instead.}%
\else
\language=\csname l@#1\endcsname
\fi
#2}}

\bibitem{schoukens2019nonlinear}
J.~Schoukens and L.~Ljung, ``Nonlinear system identification: A user-oriented
  roadmap,'' \emph{IEEE Cont. Sys. Mag.}, vol.~39, no.~6, pp. 28--99, 2019.

\bibitem{bombois2006least}
X.~Bombois, G.~Scorletti, M.~Gevers, P.~M. Van~den Hof, and R.~Hildebrand,
  ``Least costly identification experiment for control,'' \emph{Automatica},
  vol.~42, no.~10, pp. 1651--1662, 2006.

\bibitem{tanaskovic2014optimal}
M.~Tanaskovic, L.~Fagiano, and M.~Morari, ``On the optimal worst-case
  experiment design for constrained linear systems,'' \emph{Automatica},
  vol.~50, no.~12, pp. 3291--3298, 2014.

\bibitem{annergren2017application}
M.~Annergren, C.~A. Larsson, H.~Hjalmarsson, X.~Bombois, and B.~Wahlberg,
  ``Application-oriented input design in system identification: Optimal input
  design for control [applications of control],'' \emph{IEEE Cont. Sys. Mag.},
  vol.~37, no.~2, pp. 31--56, 2017.

\bibitem{bombois2021robust}
X.~Bombois, F.~Morelli, H.~Hjalmarsson, L.~Bako, and K.~Colin, ``Robust optimal
  identification experiment design for multisine excitation,''
  \emph{Automatica}, vol. 125, p. 109431, 2021.

\bibitem{pronzato2008optimal}
L.~Pronzato, ``Optimal experimental design and some related control problems,''
  \emph{Automatica}, vol.~44, no.~2, pp. 303--325, 2008.

\bibitem{hjalmarsson2009system}
H.~Hjalmarsson, ``System identification of complex and structured systems,''
  \emph{Eu. Jour. of Cont.}, vol.~15, no. 3-4, pp. 275--310, 2009.

\bibitem{de2016d}
A.~De~Cock, M.~Gevers, and J.~Schoukens, ``D-optimal input design for nonlinear
  {FIR}-type systems: A dispersion-based approach,'' \emph{Automatica},
  vol.~73, pp. 88--100, 2016.

\bibitem{valenzuela2015graph}
P.~E. Valenzuela, C.~R. Rojas, and H.~Hjalmarsson, ``A graph theoretical
  approach to input design for identification of nonlinear dynamical models,''
  \emph{Automatica}, vol.~51, pp. 233--242, 2015.

\bibitem{parsa2023coherence}
J.~Parsa, C.~R. Rojas, and H.~Hjalmarsson, ``Coherence-based input design for
  nonlinear systems,'' \emph{IEEE Cont. Sys. Let.}, vol.~7, pp. 2934--2939,
  2023.

\bibitem{heinz2017iterative}
T.~O. Heinz and O.~Nelles, ``Iterative excitation signal design for nonlinear
  dynamic black-box models,'' \emph{Proc. Comp. Sci.}, vol. 112, pp.
  1054--1061, 2017.

\bibitem{smits2024space}
V.~Smits and O.~Nelles, ``Space-filling optimized excitation signals for
  nonlinear system identification of dynamic processes of a diesel engine,''
  \emph{Cont. Eng. Prac.}, vol. 144, p. 105821, 2024.

\bibitem{peter2019fast}
T.~J. Peter and O.~Nelles, ``Fast and simple dataset selection for machine
  learning,'' \emph{Automatis.}, vol.~67, no.~10, pp. 833--842, 2019.

\bibitem{KISS2024562}
M.~Kiss, R.~Tóth, and M.~Schoukens, ``Space-filling input design for nonlinear
  state-space identification,'' in \emph{Proc. of the 20th IFAC Symp. on Sys.
  Id.}, 2024, pp. 562--567.

\bibitem{vater2024differentiable}
H.~Vater and O.~Wallscheid, ``Differentiable model predictive excitation:
  Generating optimal data sets for learning of dynamical system models,''
  \emph{Aut. Prep.}, 2024.

\bibitem{herkersdorf2024optimized}
M.~H. Herkersdorf, T.~K{\"o}sters, and O.~Nelles, ``Optimized excitation signal
  tailored to pertinent dynamic process characteristics,''
  \emph{IFAC-PapersOnLine}, vol.~58, no.~28, pp. 1049--1054, 2024.

\bibitem{smits2021genetic}
V.~Smits and O.~Nelles, ``Genetic optimization of excitation signals for
  nonlinear dynamic system identification.'' in \emph{Proc. of the Int. Conf.
  on Inf. in Cont., Aut. and Rob.}, 2021, pp. 138--145.

\bibitem{smits2022space}
------, ``Space-filling optimization of excitation signals for nonlinear system
  identification,'' in \emph{Proc. of the International Conference on
  Informatics in Control, Automation and Robotics}, 2022, pp. 255--262.

\bibitem{buisson2020actively}
M.~Buisson-Fenet, F.~Solowjow, and S.~Trimpe, ``Actively learning {G}aussian
  process dynamics,'' in \emph{Proc. of the Learn. for Dyn. and Cont.
  Conf.}\hskip 1em plus 0.5em minus 0.4em\relax PMLR, 2020, pp. 5--15.

\bibitem{treven2023optimistic}
L.~Treven, C.~Sancaktar, S.~Blaes, S.~Coros, and A.~Krause, ``Optimistic active
  exploration of dynamical systems,'' \emph{Adv. in Neur. Inf. Proc. Sys.},
  vol.~36, pp. 38\,122--38\,153, 2023.

\bibitem{nijmeijer1990nonlinear}
H.~Nijmeijer and A.~Van~der Schaft, \emph{Nonlinear dynamical control
  systems}.\hskip 1em plus 0.5em minus 0.4em\relax Springer, 1990, vol. 175.

\bibitem{rasmussen2003gaussian}
C.~E. Rasmussen, ``Gaussian processes in machine learning,'' in \emph{Advanced
  Lectures on Machine Learning}.\hskip 1em plus 0.5em minus 0.4em\relax Berlin,
  Germany: Springer, 2004, pp. 63--71.

\bibitem{johnson_minimax_1990}
M.~Johnson, L.~Moore, and D.~Ylvisaker, ``\BIBforeignlanguage{en}{Minimax and
  maximin distance designs},'' \emph{\BIBforeignlanguage{en}{Jour. of Stat.
  Plan. and Inf.}}, vol.~26, no.~2, pp. 131--148, 1990.

\end{thebibliography}

\end{document}